# A continuously tunable acoustic metasurface for transmitted wavefront manipulation

Sheng-Dong Zhao[1,2], Yue-Sheng Wang[1,*], and Chuanzeng Zhang[2,*]

[1]Institute of Engineering Mechanics, Beijing Jiaotong University, Beijing 100044, China

[2]Department of Civil Engineering, University of Siegen, Siegen 57068, Germany

**Abstract:** Previously reported acoustic metasurfaces that consist of fixed channels, are untunable to meet the broadband requirement and alterable functionalities. To overcome this limitation, we propose screw-and-nut mechanism of tunability and design a type of continuously tunable acoustic metasurface with unit components of helical cylinders which are screwed into a plate. The spiral channel length can be tuned continuously by the screwed depth; and then a metasurface with continuously tunable acoustic phase at independent pixels is attained. Different distributions of the unit components can shape an arbitrary metasurface profile. We also developed an approximate equivalent medium model to predict the tunability of the unit component. As an example, we present the design details of a circular tunable metasurface for three-dimensional acoustic focusing of different airborne sound sources in a wide frequency region. A sample of the metasurface is manufactured by poly lactic acid (PLA) plastic with the helical cylinders being 3D-printed. Experiments of sound focusing are performed. It is shown that the results of the equivalent medium model, the finite element simulation and the experiments are in a good agreement.

**Introduction**

Recently, manipulation of waves by metasurfaces becomes a research focus in the communities of electromagnetics and acoustics. A metasurface is a kind of artificial material layer with thickness being smaller than a wave length. It can modulate wavefront relying on the gradually accumulated phase changes along the metasurface. The shape of the modulated wavefront is governed by generalized Snell's law[1,2]; and needed phase profile is achieved by careful design of gradually varying sub-wavelength microstructures of the metasurface[3-7]. This concept was first proposed in the electromagnetic field[1], and was later extended to acoustics. Metasurfaces show prospective applications in many fields, e.g. design of innovative devices, such as optical vertex[1], light bending devices[3], light axicon[8], etc. due to their novel physical properties[1,3,8-10].

However, because there is no plasmon resonance in the acoustic field, design of an acoustic metasurface is a challenging task. In 2012, Liang and Li[11] used the curled perforations to coil up the space and proposed a new metamaterials with extreme constitutive parameters. Inspired by their study, many researchers proposed various designs of acoustic metasurfaces with space-coiling structures and demonstrated a variety of novel features[12-15]. The space-coiling metasurface show an exceptional ability in controlling acoustic waves, including self-accelerating beam generators[15], negative bulk modulus cell[16], acoustic rectifiers[17,18], optimal sound-absorbing device[19-21], acoustic rainbow trapping[22], extraordinary transmission[23,24], acoustic one-way device[25-27], negative refraction[11,28], unidirectional acoustic cloak[29,30], and acoustic focusing[31-34]. Such metasurface functionalities are based on the dispersive wavefront modulation and the extraordinary acoustic transmission which is attributed to the interplay between Fabry-Perot resonances inside the sub-wavelength-scale channel and resonances of the surface waves propagating on the plate[31,33].

---

[*] Corresponding author: yswang@bjtu.edu.cn

[*] Corresponding author: c.zhang@uni-siegen.de

It is known that the traditional metamaterial with a fixed microstructure can only operate in a fixed frequency range with fixed functions. In order to break this restriction, researchers pay more attentions on tunable metamaterials[35-39]. However research on tunable metasurfaces is limited. Very recently, by introducing the thermo-optic effect[40,41] or varactor diode[42,43], etc. into the active unit component of the metasurface, the continuously changeable optical phase at independent pixel achieves a great breakthrough. Continuously tunable metasurfaces have the ability to generate arbitrary radiation patterns which have long been pursued. For acoustic waves, Xie *et al.*[44] proposed the concept of coding acoustic metasurfaces that can combine simple logical bits to acquire sophisticated functions in the wave control. Memoli *et al.*[45] developed the notation of quantal metasurface that is assembled by replaceable metamaterial bricks. The tunable mechanisms in above mentioned studies are based on assembly of various fixed components, and therefore are inflexible and not continuously adjustable. Continuous tunability is much helpful to realize broadband muti-functions of metasurfaces.

In this paper, we propose a continuously tunable helical-structured metasurface (HSM) which is manufactured by screwing the helical cylinders into a plate. The spiral channel length can be tuned by the screwed depth into the plate. The HSM can slow down the propagating waves by introducing helical wave rotation and wavefront revolution[15]. Therefore, a tunable metasurface with a high effective refractive index and effective mass density is obtained. We particularly design a circular metasurface for three-dimensional acoustic focusing and perform the numerical simulation, develop the equivalent theoretical model and experimentally evidence the transmitted acoustic focusing phenomenon. The tunability of the metasurface regarding the acoustic source types (i.e. the plane wave source and the point source), the incident wave frequency, and the focal length is analyzed and discussed.

**Results**

**Basic idea of tunable metasurface design.** The design of the metasurface is illustrated schematically in Fig. 1. Fig. 1a shows the metasurface with distributed helical cylinders screwed into a plate. The tunability of the metasurface is based on the screw-and-nut mechanism with a tunable screwed depth, as shown in Fig. 1b. The plate is perforated with circular holes; and the inwall of each hole is cut into left-handed screw with shallow grooves (see Fig. 1d). After screwing the helical cylinder (Fig. 1c) into the hole, a spiral channel is formed between the inwall of the hole and the helical cylinder. The channel length and thus the transmitted phase can be continuously tuned by adjusting the screwed depth. So a new degree of freedom of controlling wavefronts can be attained by introducing arbitrary form of phase shift along the metasurface. The wavefront modulation is based on the Snell's law[1]. For the sake of simplicity, we suppose a one-way phase gradient along *x*-direction as shown in Fig. 1a. For an incident wave in *xz*-plane, the refraction angle generalized by Snell's law[1] is

$$\theta_t = \arcsin\left(\sin\theta_i + \frac{\lambda}{2\pi}\frac{\mathrm{d}\varphi}{\mathrm{d}x}\right) \qquad (1)$$

where $\theta_i$ is the incident angle; $\varphi$ is the transmitted phase; and $\lambda$ is the wave length, see the schematic diagram of Fig. 1a.

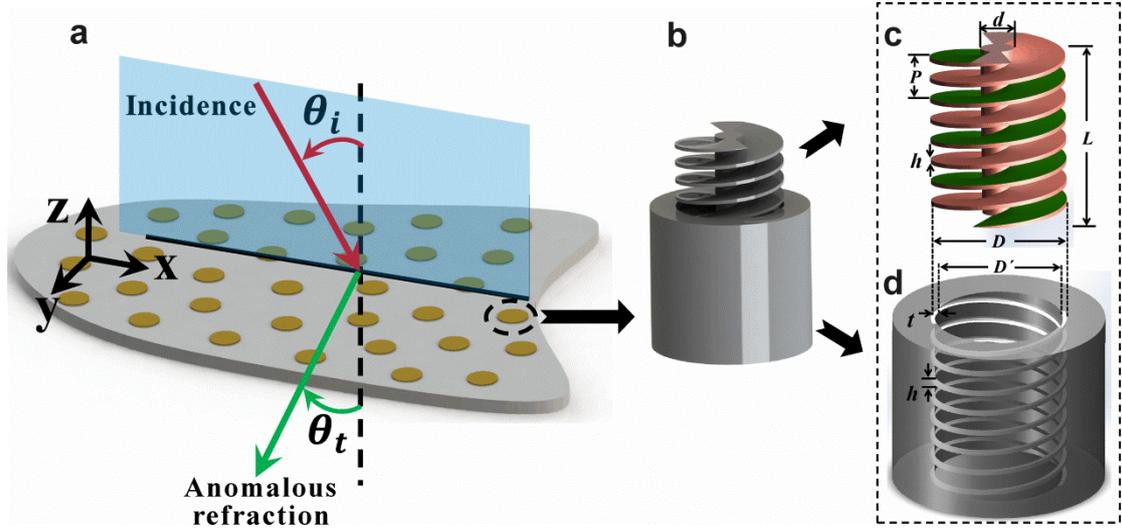

**Figure 1 | The tunable metasurface based on the screw-and-nut mechanism.** (a) The schematic diagram shows that the anomalous refraction phenomenon achieved by the tunable metasurface is governed by the generalized Snell's law. The gradient phase is tuned by the screwed depth of the helical cylinders. (b) The helical cylinder with two opposite spiraling blades connected through a central slender column. The cylinder is screwed into a block and then a tunable unit component is combined. (c) The geometric details of the helical cylinder. (d) The drilled hole with the inwall cut into shallow grooves.

One of the key steps in design of the tunable metasurface is to obtain the relationship between the phase changes and screwed depth of the helical cylinder under different frequencies with consideration of high transmission. As an example, we will next present a careful design of a circular tunable metasurface for three-dimensional acoustic focusing of different airborne sound sources in a wide frequency region.

**Design of a tunable three-dimensional focusing metasurface.** The proposed tunable HSM is composed of a perforated circular plate and helical cylinders screwed into the plate. All the components are made of poly lactic acid (PLA) plastic which can be treated as a rigid material relative to the airborne sound. Each helical cylinder is 3D-printed with two opposite spiraling blades connected through a central slender column, see Fig. 2a. Its geometry can be defined by the outer diameter $D=32\text{mm}$, the column diameter $d=8\text{mm}$, the length $L=40\text{mm}$ which is the same as the plate thickness, the blade thickness $h=1\text{mm}$, and the thread lead $P=10\text{mm}$. The cutting shallow grooves on the inwall of the drilled hole are defined by $t=1\text{mm}$ in depth and $h=1\text{mm}$ in width (see Fig. 1d). All the above geometric parameters marked in Fig. 1c, d are smaller than the wavelength of the airborne sound. The plate with the radius of $r_d=190\text{mm}$ is perforated with circular holes distributed in four circular layers as shown in Fig. 2b, c. The diameter of the hole is $D'=30\text{mm}$. The first layer is in the plate center with one hole; and the other three layers from the center to the outside contain 6, 12 and 18 holes, respectively. The distance between two adjacent layers is 50mm (i.e., $r_1=0$, $r_2=50\text{mm}$, $r_3=100\text{mm}$, $r_4=150\text{mm}$, see Fig. 2b). The helical cylinders are screwed into the holes as shown in Fig. 2c. Then a tunable spiral channel is formed between the inwall of the hole and the helical cylinder.

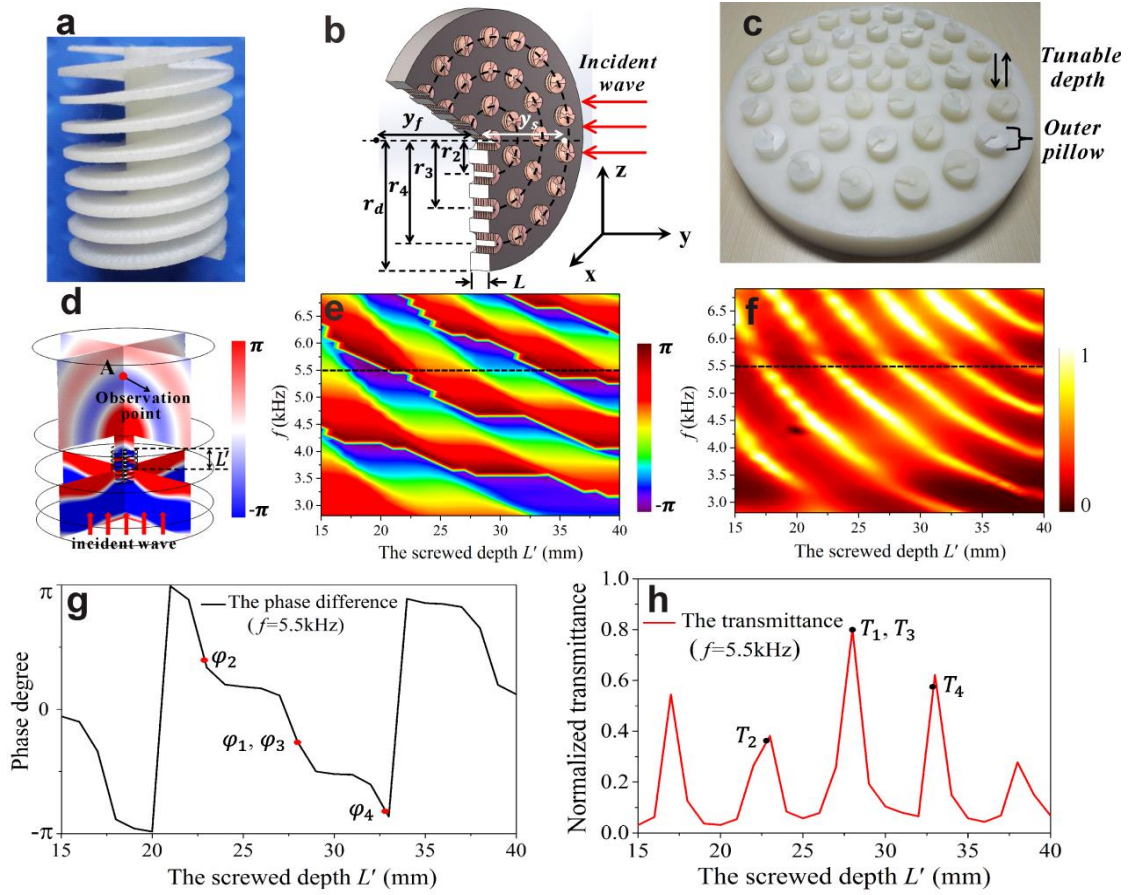

**Figure 2 | The tunable metasurface for three-dimensional acoustic focusing and the transmission property corresponding to the screwed depth.** (a) The fabricated helical cylinder based on 3D printing technique. (b) The geometric details of the plate with distributed helical cylinders screwed into the holes. (c) The sample of the tunable metasurface composed of a perforated circular plate with four layers of the helical air channels formed by the helical cylinders. (d) The transmitted wave field from the numerical simulation model with the screwed depth of $L' = 22\text{mm}$ at the frequency of 5.5kHz. (e) and (f) The nephograms of the phase and transmittance in the plane of the frequency and screwed depth. (g) and (h) The cut lines of (e) and (f) at the frequency of 5.5kHz showing the variation of the phase and transmittance with the screwed depth.

**Introduction of working principle and performance.** We consider acoustic waves which are propagating through the HSM along the helical paths. By tuning the screwed depth of the helical cylinders appropriately, we can obtain the proper radial gradient variation of the phase distribution, yielding acoustic focusing according to the generalized Snell's law. This is the basic idea of our tunable metasurface. In the present design, the tunable depth ranges from 15mm to 40mm inducing highly controllable phase shifts of the transmitted wavefront over the whole $2\pi$ range within the focusing frequency between 3.9 kHz and 6.3 kHz. Based on the flexible wave-control capability, a variety of tunable items including the broadband frequency range, the tunable focusing length $y_f$, and the alterable acoustic source (an incident plane wave source or a point source) are verified.

To realize the acoustic focusing with a particular source at a particular frequency, we should first establish the relationship between the transmitted wave phase and the screwed depth of the helical cylinders. To this end, a finite element simulation model based on the commercial software COMSOL Multiphysics is developed, see Fig. 2d.

The calculated phase shift and the normalized transmittance of the transmitted waves at the observation point A, which has a distance of 90mm to the plate, are shown in Fig. 2e, f as functions of the frequency and the screwed depth.

By properly selecting the phase profiles in the radial direction of the metasurface plate, several fascinating wavefront engineering capabilities can be realized. For acoustic focusing, the hyperboloidal phase profile is necessary[34]. Because only the phases at the four layers of the helical cylinders can be tuned, the continuous phase profile must be converted to a discrete phase profile. For an incident plane wave, we obtain the phase difference between adjacent layers as

$$\varphi_i - \varphi_{i-1} = \frac{2\pi(d_i - d_{i-1})}{\lambda} \in [-\pi, \pi], \quad i = 2, 3, 4 \qquad (2)$$

where $\lambda = c_0/f$ is the wave length at the focusing frequency $f$ with $c_0 = 342 \text{ms}^{-1}$ being the airborne sound speed and $d_i = (y_f^2 + r_i^2)^{1/2}$ is the wave path from the $i$th layer to the focal point, see Fig. 2b. For a point wave source, the associated phase difference is given by

$$\varphi_i - \varphi_{i-1} = \frac{2\pi(d_i - d_{i-1} + d_i' - d_{i-1}')}{\lambda} \in [-\pi, \pi], \quad i = 2, 3, 4 \qquad (3)$$

where $d_i' = (y_s^2 + r_i^2)^{1/2}$ is the wave path from the $i$th layer to the source point with $y_s$ being the distance between the point source and the hyperlens. Within the $2\pi$ controllable phase shifts of the transmitted waves, if the phase degree of the $i$th layer ($\varphi_i$) is determined, other phase degrees are thereupon determined by following the Eq. (2) or (3). It should be noted here that the result obtained in this way is not unique. Among the numerous potential tunable results, we should find the optimal one to achieve a high transmittance. So we should always try to make as many helical cylinders as possible to have a high transmittance when we determine their screwed depth.

As an example, we consider an incident plane wave with the frequency $f = 5.5\text{kHz}$. For the focusing length $y_f = 50\text{mm}$, we can find that the phase differences between the first layer to the others are $\varphi_2 - \varphi_1 = 2.1$, $\varphi_3 - \varphi_1 = -0.027$ and $\varphi_4 - \varphi_1 = -1.63$ according to Eq. (2). The cut lines of the phase and transmittance nephogram (see the dashed lines in Fig. 2e, f, respectively) at the frequency of 5.5 kHz are plotted in Fig. 2g, h. Assuming that $\varphi_3 \approx \varphi_1 = -0.81$ then we can obtain $\varphi_2 = 1.29$ and $\varphi_4 = -2.44$ as shown in Fig 2g. The corresponding transmittance $T_1$, $T_2$, $T_3$ and $T_4$ are plotted in Fig. 2h, which are all located around the resonance peaks. Finally we get the optimized screwed depths $L'$ of the four layers (from the center to the outside) as: 28mm, 22.8mm, 28mm and 32.8mm. Fig. 3a shows the schematic diagram of the plane wave focusing model. The calculated normalized pressure intensity field $(P/P_0)^2$ is presented in Fig. 3b. It is seen that the selected screwed depths of the helical cylinders can indeed guarantee a satisfactory acoustic focusing at the frequency of $f = 5.5\text{kHz}$. Following the same process as described above, the optimized screwed depth corresponding to different frequencies and focusing lengths for both plane wave and point sources are obtained, and the results are given in Table S1 of the Supporting Information. These data may be used to tune the focus of the designed HSM.

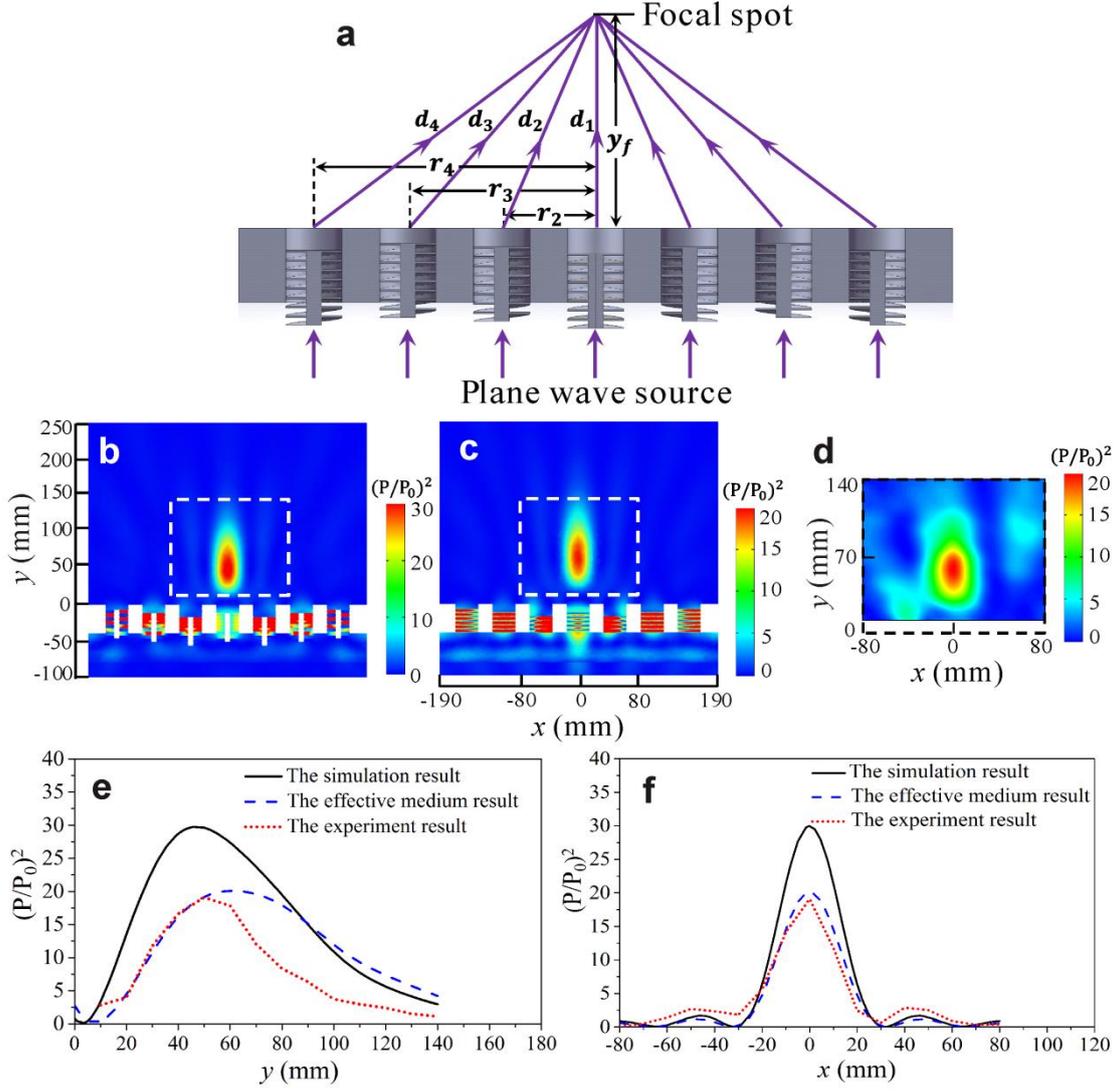

**Figure 3 | The focusing results for an incident plane wave source at the frequency of 5.5kHz with the focusing length $y_f = 50$ mm.** (a) The schematic diagram of the plane wave focusing model. (b) The simulation results. (c) The effective medium model results. (d) The experimental results. (e) and (f) The longitudinal and transverse pressure intensity fields at the focal spot along the y- and x-directions. The pressure intensity is normalized by the incident plane wave amplitude $P_0$.

**Comparison of the numerical simulation, approximate effective medium model and experimental results.** Instead of the above detailed numerical simulation, an approximate effective medium model is developed to design the tuning of the acoustic focusing. In our model, the helical cylinder with the air in the spiral channel is replaced by a cylindrical channel with an equivalent homogeneous medium which has the same size as the screwed part (with the diameter $D' = 30$ mm and length $L'$). It is noted here that the pillow left out of the plate should be considered in the model. The effective density $\rho_{eff}$ (in kg/m$^3$) and the refractive index $n_{eff}$ are calculated as functions of the screwed depth $L'$. For more details of the model, we refer to the Supplementary Note 1. The results are as follows:

$$n_{eff}(L') = 3.37 \times 10^{-5}(L')^3 - 3.47 \times 10^{-3}(L')^2 + 0.129L' + 3.85, \quad (4)$$

$$\rho_{eff}(L') = 3.07 \times 10^{-4}(L')^3 - 3.17 \times 10^{-2}(L')^2 + 1.18L' + 35.11, \quad (5)$$

from which we can easily obtain the effective parameters associated with the screwed depth. The acoustic focusing field based on the approximate effective medium model is plotted in Fig. 3c. The experimentally measured pressure intensity field is shown in Fig. 3d. Fig. 3b-d all show a sharp focal spot that is about more than four times of the pressure amplitude of the incident plane wave. For a more quantitative comparison, the normalized longitudinal and transverse pressure intensity fields at the focal spot ($y = y_f$) along the *y*- and *x*-directions corresponding to Fig. 3b-d are shown in Fig. 3e, f. All results from the numerical simulation, effective medium model and experimental measurement are in good agreement except that the intensities obtained from the effective medium model and the experimental measurement are a little lower than the numerical simulation result. Presumably, this discrepancy is caused by the fact that an energy loss always exists due to the air viscosity in the experiment, and the big entrance part (Supplementary Fig. 1) induces a low estimation of the resonant transmission peaks by the approximate effective medium model. The simulation and experimental results for some other selected values of the frequency (4.1kHz, 5.1kHz and 5.9kHz) and focusing lengths (50mm and 100mm) are plotted in Supplementary Fig. 3. These results show that a broadband acoustic focusing can be achieved by the tunable HSM for an incident plane wave.

  We next discuss the focusing of a point wave source with the source length of $y_s = 150\text{mm}$ (Fig. 4a). In this example, the focusing frequency $f = 5.5\text{kHz}$ is also assumed but the focusing length is $y_f = 100\text{mm}$. The phase degrees are thereupon determined by following Eq. (3), and the verified phase differences between the first layer to the others are $\varphi_2 - \varphi_1 = 2.01$, $\varphi_3 - \varphi_1 = 0.97$ and $\varphi_4 - \varphi_1 = 1.85$, respectively. Then the optimized screwed depths of the four layers in this example are selected from the Table S1 of the Supporting Information as 32mm, 27.6mm, 29mm and 28mm, respectively. The results of the numerical simulation, the effective medium model and the experiment are shown in Fig. 4b-d. The normalized pressure intensity fields of $(P/P_s)^2$ in the *y*- and *x*-directions across the focusing point show that the focusing length is indeed around 100mm (Fig. 4e, f). Similar to the plane wave incidence, the numerical simulation and experimental results for some other selected values of the frequency (4.3kHz, 5.1kHz and 6.1kHz) and focusing lengths (50mm and 100mm) are plotted in Supplementary Fig. 4 of the Supporting Information. The results sufficiently illustrate that the tunable HSM is effective for the broadband tunable acoustic focusing.

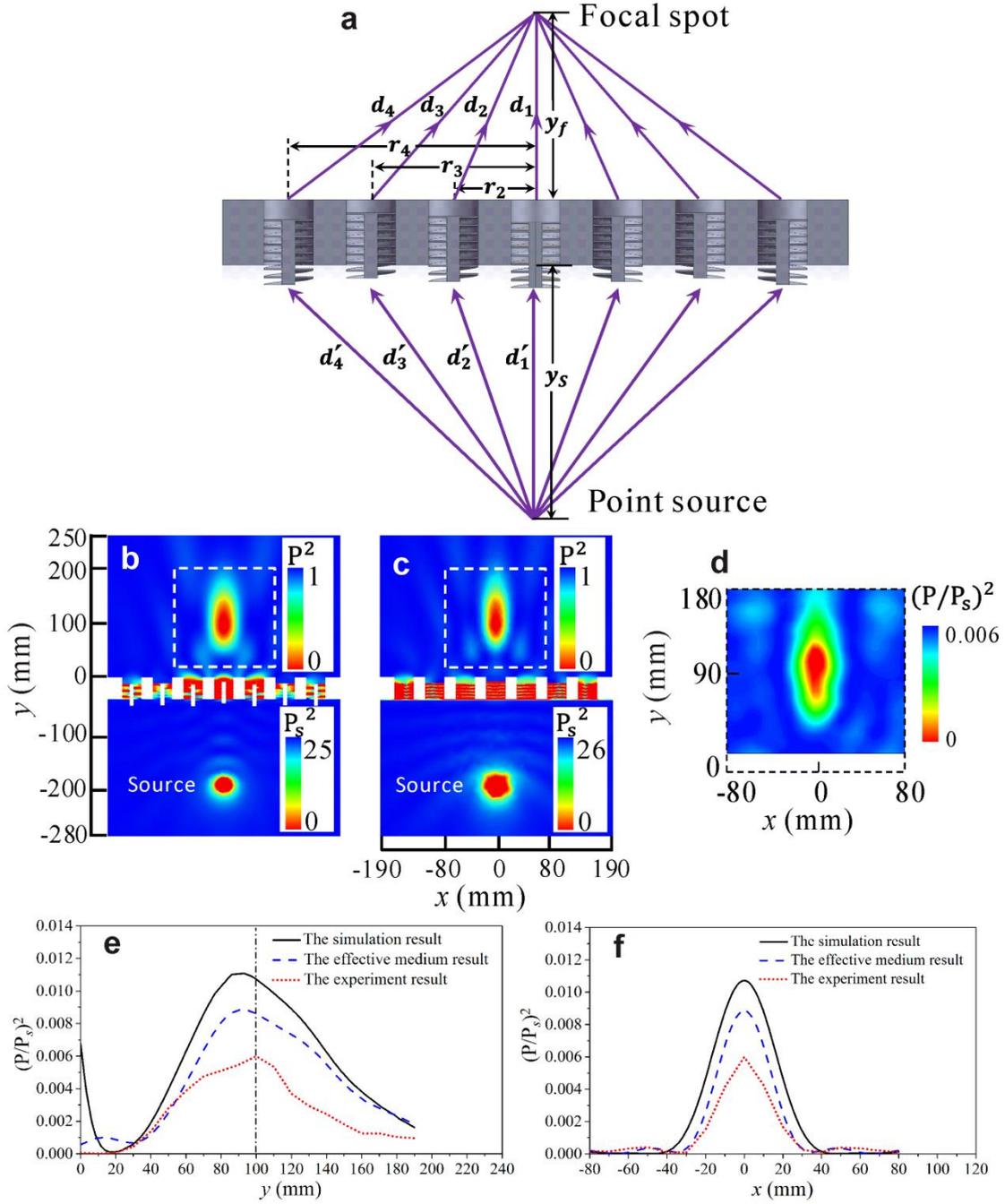

**Figure 4 | The focusing results for a point source at the frequency of 5.5kHz with the source length $y_s = 150$mm and focusing length $y_f = 100$mm.** (a) The schematic diagram of the point source focusing model. (b) The simulation results. (c) The effective medium model results. (d) The experimental results. (e) and (f) The longitudinal and transverse pressure intensity fields at the focal spot along the y- and x-directions. The pressure intensity is normalized by the pressure amplitude of a semi-spherical surface source $P_s$ (with radius of 12.7mm). The detailed definition and calculation are given in Supplementary Note 3.

**Collection and analysis of the data of the focusing quality.** To evaluate the focusing quality and the broadband tunability within the frequency range of 3.9~6.3kHz, we present the numerical simulation results including the focusing lengths, the focusing amplitudes and the focusing resolutions in Fig. 5. The data are collected into four groups corresponding to the two kinds of acoustic sources (plane wave source and point source) and the two preset focusing lengths (50mm and 100mm). The optimized values of the screwed depth are taken from Table S1. It is seen from Fig. 5a that the four groups of the data are located around their preset focusing positions of $y_f = 50$mm and $y_f = 100$mm with a maximum error of 20%. The four groups of the focusing amplitudes are shown in Fig. 5b in two scales. The pressure amplitudes of the focusing point with respect to the point source are around 12%, and the focusing amplitudes for the plane wave source are around 4.5 times of the incident plane wave amplitude. Fig. 5c, d show the longitudinal and transverse spatial resolutions which are defined as the full widths at the half maximum of the transmission peak measured crossing the focusing point along the $y$- and $x$-directions, respectively. The results demonstrate that the transverse resolution is higher than the longitudinal resolution, which means that the focusing point is not a circle. In particular, the near field focusing (50mm) acquires a higher resolution than the far focusing (100mm). The transverse resolution does not break but is close to the diffraction limit of $0.5\lambda$.

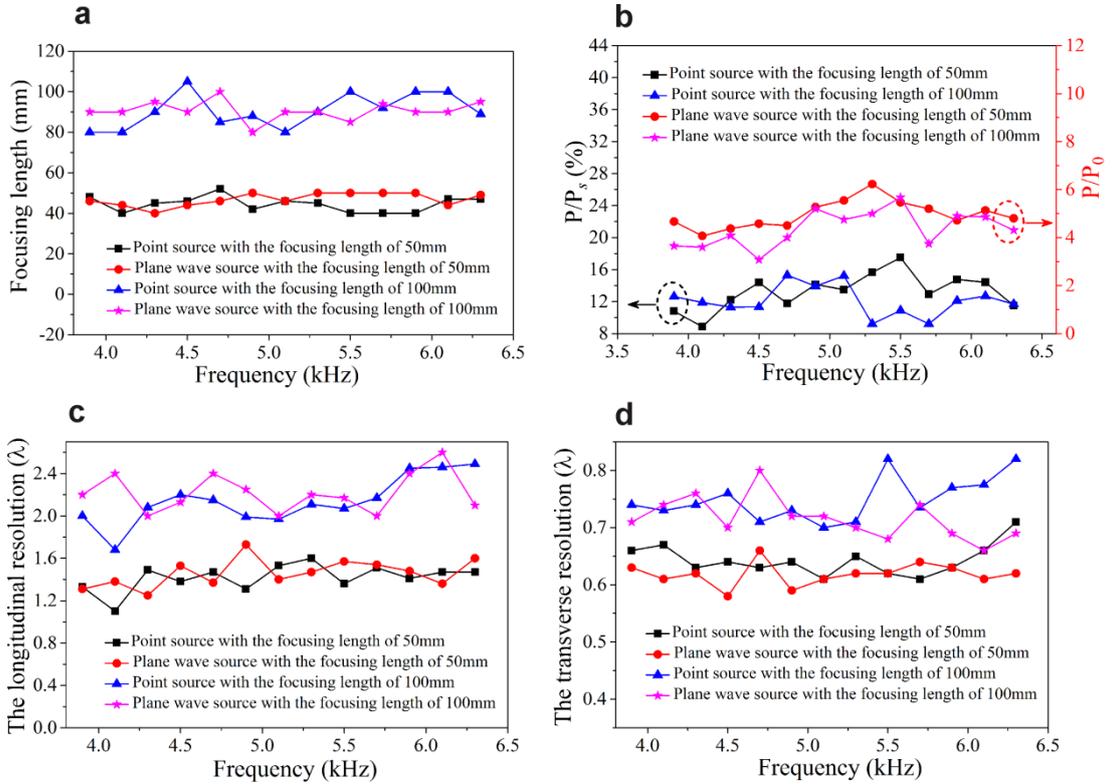

**Figure 5 │ All the details of the simulation results for the broadband tunable focusing effect.** The numerical simulation results of the focusing lengths (a), the focusing amplitudes (b) and the longitudinal (c) and transverse (d) focusing resolutions for the plane wave and point sources with the focusing length of 50mm and 100mm within the frequency range of 3.9kHz~6.3kHz.

**Discussion**

In summary, we designed a novel tunable metasurface consisting of helical cylinders screwed into a circular plate. The screw-and-nut mechanism is used to tune the spiral channel length to obtain a flexible manipulated wavefront. Thus, a continuously tunable acoustic metasurface is achieved. The broadband tunable acoustic focusing property of the system is studied in details. Acoustic focusing with a relatively high transmittance is obtained, and the tunability is independent of the sound source type and the focusing length. Such flexible tenability is highly desirable in phase engineering applications. The proposed mechanism can improve the adjustment of the metasurface having any kind of the phase profile, such as the acoustic axicon for the Bessel beam or Airy beam, tunable carpet cloak and even a lens with tunable helical wavefronts that carry an angular momentum. Furthermore, the proposed unit components can be assembled into a three-dimensional curved surface to design a tunable curved metasurface.

**Method**
**Numerical simulations.** Throughout the paper, all wave propagation simulations are performed by using COMSOL Multiphysics 4.4 with the pressure acoustics module. The surrounding gas applied in our simulations is air. The solid metasurface is treated as a rigid material relative to the airborne sound and sound hard boundaries are used on the surface. The plane wave radiation boundary condition is adopted on the outer boundaries to eliminate the reflected waves.

**Sample fabrication and experimental setup.** Each helical cylinder was fabricated by using a 3D printer (Z500, HORI, China). The circular plate was fabricated through mechanical processing.

In the experiment, the acoustic wave was incident to the HSM from the side with outer pillars (Supplementary Fig. 5a), and the pressure intensity field was measured on the other side (Supplementary Fig. 5b). The tunable lens was surrounded by the absorbing foam and placed on the axis of the sound radiation. The loudspeaker (HS8, YAMAHA, Japan) is surrounded by the absorbing foam box with the thickness of 50mm (Supplementary Fig. 5c). For the incident plane wave source, the loudspeaker was placed on the lens axis with the distance of 1300mm (marked by 1 in Supplementary Fig. 5c) which is far enough to generate a plane wave. For the point source the loudspeaker was placed at the source point with the distance of 150mm to the lens (marked by 2 in Supplementary Fig. 5c). In the experiment, we first used the software of Audacity to generate a single frequency signal, and then the signal was input to the sound card (AudioBox22VST, PreSonus, USA) and played out by the loudspeaker. On the other side of the lens, the pressure intensity field was scanned in the axial plane area in a rectangular region of $160 \times 140 \text{mm}^2$ for the focusing length of 50mm, and $160 \times 180 \text{mm}^2$ for the focusing length of 100mm. The scanned area was separated from the right lens edge by 10mm. The scanned signal was acquired and analyzed by Audacity.

**Data availability.** The data that support the findings of this study are available from the corresponding author on reasonable request.

**Reference**

1. Yu, N., *et al.* Light propagation with phase discontinuities: generalized laws of reflection and refraction. *Science* **334**, 333-337 (2011).
2. Chen, H.T., Taylor, A.J. & Yu, N. A review of metasurfaces: physics and applications. *Rep. Prog. Phys.* **79**, 076401 (2016).
3. Ni, X., Emani, N.K., Kildishev, A.V., Boltasseva, A. & Shalaev, V.M. Broadband light bending with plasmonic nanoantennas. *Science* **335**, 427 (2012).
4. Sun, S.L., *et al.* Gradient-index meta-surfaces as a bridge linking propagating waves and surface waves. *Nat. Mater.* **11**, 426-431 (2012).
5. Sun, S., *et al.* High-efficiency broadband anomalous reflection by gradient

meta-surfaces. *Nano. Lett.* **12**, 6223-6229 (2012).

6. Arbabi, A., Horie, Y., Bagheri, M. & Faraon, A. Dielectric metasurfaces for complete control of phase and polarization with subwavelength spatial resolution and high transmission. *Nat. Nanotechnol.* **10**, 937-943 (2015).
7. Wolf, O., *et al.* Phased-array sources based on nonlinear metamaterial nanocavities. *Nat. Commun.* **6**, 7667 (2015).
8. Lin, D., Fan, P., Hasman, E. & Brongersma, M.L. Dielectric gradient metasurface optical elements. *Science* **345**, 298-302 (2014).
9. Li, G., *et al.* Continuous control of the nonlinearity phase for harmonic generations. *Nat. Mater.* **14**, 607-612 (2015).
10. Raman, A.P., Anoma, M.A., Zhu, L., Rephaeli, E. & Fan, S. Passive radiative cooling below ambient air temperature under direct sunlight. *Nature* **515**, 540-544 (2014).
11. Liang, Z. & Li, J. Extreme acoustic metamaterial by coiling up space. *Phys. Rev. Lett.* **108**, 114301 (2012).
12. Xie, Y., Popa, B.I., Zigoneanu, L. & Cummer, S.A. Measurement of a broadband negative index with space-coiling acoustic metamaterials. *Phys. Rev. Lett.* **110**, 175501 (2013).
13. Li, Y., *et al.* Acoustic focusing by coiling up space. *Appl. Phys. Lett.* **101**, 233508 (2012).
14. Li, Y., *et al.* Three-dimensional Ultrathin Planar Lenses by Acoustic Metamaterials. *Sci. Rep.* **4**, 6830 (2014).
15. Zhu, X.F., *et al.* Implementation of dispersion-free slow acoustic wave propagation and phase engineering with helical-structured metamaterials. *Nat. Commun.* **7**, 11731 (2016).
16. Cheng, Y., *et al.* Ultra-sparse metasurface for high reflection of low-frequency sound based on artificial Mie resonances. *Nat. Mater.* **14**, 1013-1019 (2015).
17. Liang, B., Yuan, B. & Cheng, J.C. Acoustic diode: rectification of acoustic energy flux in one-dimensional systems. *Phys. Rev. Lett.* **103**, 104301 (2009).
18. Liang, B., Guo, X.S., Tu, J., Zhang, D. & Cheng, J.C. An acoustic rectifier. *Nat. Mater.* **9**, 989-992 (2010).
19. Yang, M., Chen, S.Y., Fuab, C.X. & Sheng, P. Optimal sound-absorbing structures. *Mater. Horiz.* **4**, 673-680 (2017).
20. Li, Y. & Assouar, B.M. Acoustic metasurface-based perfect absorber with deep subwavelength thickness. *Appl. Phys. Lett.* **108**, 063502 (2016).
21. Ma, G., Yang, M., Xiao, S., Yang, Z. & Sheng, P. Acoustic metasurface with hybrid resonances. *Nat. Mater.* **13**, 873-878 (2014).
22. Ni, X., *et al.* Acoustic rainbow trapping by coiling up space. *Sci. Rep.* **4**, 7038 (2014).
23. Li, Y., Liang, B., Zou, X.Y. & Cheng, J.C. Extraordinary acoustic transmission through ultrathin acoustic metamaterials by coiling up space. *Appl. Phys. Lett.* **103**, 063509 (2013).
24. Li, Y., Liang, B., Gu, Z.M., Zou, X.Y. & Cheng, J.C. Unidirectional acoustic transmission through a prism with near-zero refractive index. *Appl. Phys. Lett.*

**103**, 053505 (2013).

25. Jiang, X., *et al.* Acoustic one-way metasurfaces: Asymmetric Phase Modulation of Sound by Subwavelength Layer. *Sci. Rep.* **6**, 28023 (2016).
26. Zhu, Y.F., Zou, X.Y., Liang, B. & Cheng, J.C. Acoustic one-way open tunnel by using metasurface. *Appl. Phys. Lett.* **107**, 113501 (2015).
27. Li, Y., *et al.* Tunable Asymmetric Transmission via Lossy Acoustic Metasurfaces. *Phys. Rev. Lett.* **119**, 035501 (2017).
28. Zhu, H.F. & Semperlotti, F. Anomalous Refraction of Acoustic Guided Waves in Solids with Geometrically Tapered Metasurfaces. *Phys. Rev. Lett.* **117**, 034302 (2016).
29. Wang, X.P., Wan, L.L., Chen, T.N., Song, A.L. & Wang, F. Broadband unidirectional acoustic cloak based on phase gradient metasurfaces with two flat acoustic lenses. *J. Appl. Phys.* **120**, 014902 (2016).
30. Wang, X., Mao, D.X. & Li, Y. Broadband acoustic skin cloak based on spiral metasurfaces. *Sci. Rep.* **7**, 11604 (2017).
31. Al Jahdali, R. & Wu, Y. High transmission acoustic focusing by impedance-matched acoustic meta-surfaces. *Appl. Phys. Lett.* **108**, 031902 (2016).
32. Yuan, B.G., Cheng, Y. & Liu, X.J. Conversion of sound radiation pattern via gradient acoustic metasurface with space-coiling structure. *Appl. Phys. Express* **8**, 027301 (2015).
33. Moleron, M., Serra-Garcia, M. & Daraio, C. Acoustic Fresnel lenses with extraordinary transmission. *Appl. Phys. Lett.* **105**, 114109 (2014).
34. Li, Y., Liang, B., Gu, Z.M., Zou, X.Y. & Cheng, J.C. Reflected wavefront manipulation based on ultrathin planar acoustic metasurfaces. *Sci. Rep.* **3**, 2546 (2013).
35. Wang, Z.W., Zhang, Q., Zhang, K. & Hu, G.K. Tunable Digital Metamaterial for Broadband Vibration Isolation at Low Frequency. *Adv. Mater.* **28**, 9857 (2016).
36. Li, F.M., Zhang, C.Z. & Liu, C.C. Active tuning of vibration and wave propagation in elastic beams with periodically placed piezoelectric actuator/sensor pairs. *J. Sound. Vib.* **393**, 14-29 (2017).
37. Wang, P., Casadei, F., Shan, S.C., Weaver, J.C. & Bertoldi, K. Harnessing Buckling to Design Tunable Locally Resonant Acoustic Metamaterials. *Phys. Rev. Lett.* **113**, 014301 (2014).
38. Zhang, H.K., Zhou, X.M. & Hu, G.K. Shape-adaptable hyperlens for acoustic magnifying imaging. *Appl. Phys. Lett.* **109**, 224103 (2016).
39. Zhang, Q., Zhang, K. & Hu, G.K. Smart three-dimensional lightweight structure triggered from a thin composite sheet via 3D printing technique. *Sci. Rep.* **6**, 22431 (2016).
40. Sun, J., Timurdogan, E., Yaacobi, A., Hosseini, E.S. & Watts, M.R. Large-scale nanophotonic phased array. *Nature* **493**, 195-199 (2013).
41. DeRose, C.T., *et al.* Electronically controlled optical beam-steering by an active phased array of metallic nanoantennas. *Opt. Express* **21**, 5198-5208 (2013).
42. Zhu, B.O., *et al.* Dynamic control of electromagnetic wave propagation with the equivalent principle inspired tunable metasurface. *Sci. Rep.* **4**, 4971 (2014).

43. Zhu, B.O., Zhao, J.M. & Feng, Y.J. Active impedance metasurface with full 360 degrees reflection phase tuning. *Sci. Rep.* **3**, 3059 (2013).
44. Xie, B.Y._, et al._ Coding Acoustic Metasurfaces. *Adv. Mater.* **29**, 1603507 (2017).
45. Memoli, G._, et al._ Metamaterial bricks and quantization of meta-surfaces. *Nat. Commun.* **8**, 14608 (2017).

**Acknowledgements**
This work is supported by the National Natural Science Foundation of China (Grant No. 11532001) and Joint Sino-German Research Project (Grant No. GZ 1355).

**Author Contributions**
S. D. Zhao performed the numerical simulations and drafted the manuscript. Y. S. Wang and C. Zhang designed the experiment and supervised the study. S. D. Zhao performed the measurement and the data processing Y. S. Wang and C. Zhang conceived and led the project and contributed to the writing and revision. All authors contributed to the analysis and discussions.

**Author Information** The authors declare no competing financial interests. Correspondence and requests for materials should be addressed to Y. S. Wang (yswang@bjtu.edu.cn) and C. Zhang (c.zhang@uni-siegen.de).

# Supplementary Information

# A continuously tunable acoustic metasurface for transmitted wavefront manipulation

Sheng-Dong Zhao[1,2], Yue-Sheng Wang[1,*], and Chuanzeng Zhang[2,*]

[1]Institute of Engineering Mechanics, Beijing Jiaotong University, Beijing 100044, China

[2]Department of Civil Engineering, University of Siegen, Siegen 57068, Germany

[*] Corresponding author: yswang@bjtu.edu.cn

[*] Corresponding author: c.zhang@uni-siegen.de

# Supplementary Note 1. Modified effective medium model

This section presents a modified effective medium model considering the outer pillow of the screwed cylinder.

According to the effective medium model, the transmission of a one-dimensional acoustic wave propagating through a homogeneous medium of length $L'$ can be written as[1]:

$$T = \frac{4}{4\cos^2(k_0 n_{eff} L') + \left(\dfrac{\rho_{eff}}{\rho_0 n_{eff}} + \dfrac{\rho_0 n_{eff}}{\rho_{eff}}\right)^2 \sin^2(k_0 n_{eff} L')}, \tag{S1}$$

where $\rho_0 = 1.21 \text{kgm}^{-3}$ is the mass density of the air; $k_0$ is the wave propagation constant; and $\rho_{eff}$ and $n_{eff}$ are the effective density and refractive index, respectively. It can be seen from Eq. (S1) that the resonant peaks will appear when $k_0 n_{eff} L' = m\pi$ with $m$ being an integer. This is the well-known Fabry-Pérot resonant condition that defines the resonant frequencies as $f = m\dfrac{c_0}{2L'}$ where $c_0 = 342 \text{ms}^{-1}$ is the airborne sound speed. Based on the resonant peaks in Eq. (S1) and taking $m=1$, we can retrieve the effective refractive index as

$$n_{eff} = \frac{c_0}{2 f_1 L'}, \tag{S2}$$

where $f_1$ is the frequency of the first nonzero resonance peak.

When $k_0 n_{eff} L' = (2m-1)\pi/2$, Eq. (S1) yields the minimum transmittance as

$$T_{min} = \frac{4}{\left(\dfrac{\rho_{eff}}{\rho_0 n_{eff}} + \dfrac{\rho_0 n_{eff}}{\rho_{eff}}\right)^2}. \tag{S3}$$

From Eq. (S3), we can retrieve the effective mass density as

$$\rho_{eff} = \frac{\rho_0 n_{eff}}{\sqrt{T_{min}}} + \rho_0 n_{eff} \sqrt{\frac{1}{T_{min}} - 1}. \tag{S4}$$

From Eqs. (S2) and (S4) it can be seen that $f_1$ and $T_{min}$ are the two key parameters to be determined in the effective medium model. For the present tunable HSM, a pillow with the depth of $L - L'$ will be left out of the plate which can be an obstacle and should be taken into account when the effective medium model is used. In order to include the effect of the pillow, we developed a modified model which has a wider entrance part with the diameter $D_0 = 50\text{mm}$ (we have verified numerically that a larger size of $D_0$ has a negligible effect on the results) than the original model, as shown in Supplementary Fig. 1. Supplementary Fig. 1a is the idealized model without the outer pillow. Supplementary Fig. 1b is the model with a wider entrance part. Supplementary Fig. 1c is the model with a wider entrance part and an outer pillow. Supplementary Fig. 1d is the modified effective medium model corresponding to model S1c. Without loss of generality, we consider a unit-cell with a helical cylinder screwed depth of $L' = 22\text{mm}$. Supplementary Fig. 1a-d show the acoustic transmission fields of the four models at the frequency of 5.74 kHz, and the corresponding transmission ratios are plotted in Supplementary Fig. 1e where a rectangular box highlights the focusing frequency range. We can find that the peak frequencies have almost no change when a wider entrance part is introduced, see the black solid and dashed lines in Supplementary Fig. 1e. However, the outer pillow will obviously influence the resonant frequencies including the fundamental resonant peak frequency $f_1$, see the lines marked with b and c in Supplementary Fig. 1e. Nevertheless, the outer pillow has a little impact on the transmission ratio at the resonance valley which is marked by an ellipse in Supplementary Fig. 1e. So the outer pillow just affects one of two key parameters, i.e. the fundamental resonance frequency $f_1$, which will be modified and denoted as $f_m$. The other parameter $T_{min}$ in the idealized model of Supplementary Fig. 1a is still valid in the modified effective medium model when the outer-pillow effect is considered. The resonance peaks and lowest transmittance of the modified effective medium model should be consistent with the model of Supplementary Fig. 1c in the focusing frequency range marked by the rectangular box in Supplementary Fig. 1e. The modified frequency $f_m$ is retrieved from the *i*th resonance peak located in the focusing frequency range (3.9k~6.3k). We assume that the *i*th resonant frequency $f_i$ is the nearest one to the central frequency (5.1 kHz) of the focusing frequency range, then we have

$$f_m = \frac{f_i}{i}. \tag{S5}$$

Therefore the modified effective refractive index can be written as

$$n_{eff} = \frac{c_0 i}{2 f_i L'}. \tag{S6}$$

The effective mass density keeps the same form as Eq. (S4). The transmittance of the actual model (S1c) and the modified effective medium model (S1d) reach their peaks at the same frequency $f_i$, which is called the coupled peak as shown by the red dash-dot and solid lines in Supplementary Fig. 1e. The two lines keep pace with each other in the focusing frequency range, and the two models of Supplementary Fig. 1c and d show the similar transmittance acoustic field.

In the particular case of $L' = 22\text{mm}$, the coupled resonance peak is $f_i = 5.74\text{kHz}$ (*i*=4), so the effective parameters can be calculated as $n_{eff} = 5.42$ and $\rho_{eff} = 49.4 \text{kgm}^{-3}$. With the screwed depth $L'$ ranging from 15mm to 40mm, the discrete effective parameters are calculated based on the modified effective medium model. In order to obtain the approximate expressions, the effective parameters are fitted by polynomials of the 3th degree, which are functions of the screwed depth. The discrete and fitted results of the effective parameters are plotted in Supplementary Fig. 2. Then, in the acoustic focusing study we can easily calculate the effective material parameters associated with the screwed depth using these approximate expressions.

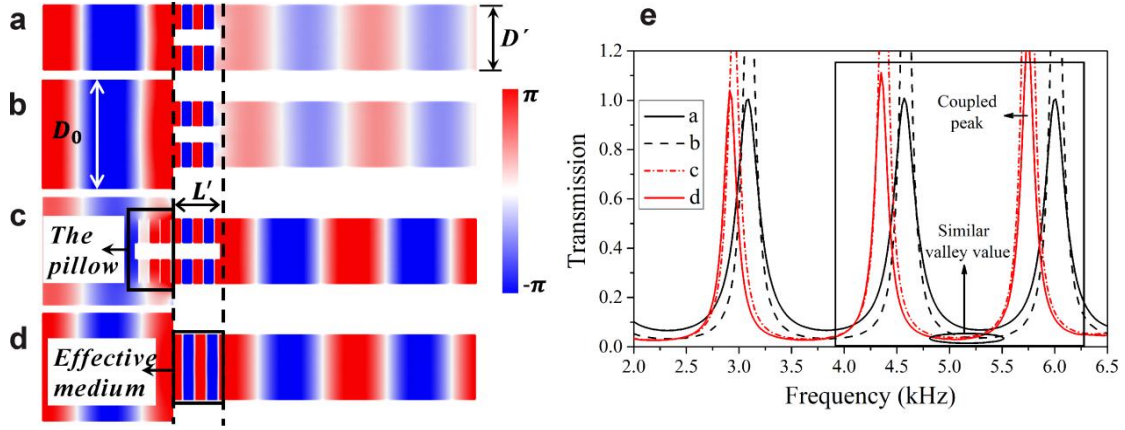

**Supplementary Figure 1** | **The acoustic transmission fields at the frequency of 5.74 kHz for the four models:** the idealized model without the outer pillow (a), the model with a wider entrance part (b), the model with an outer pillow (c), and the effective medium model (d) corresponding to the model (c). The transmission ratios of the four models are plotted in (e).

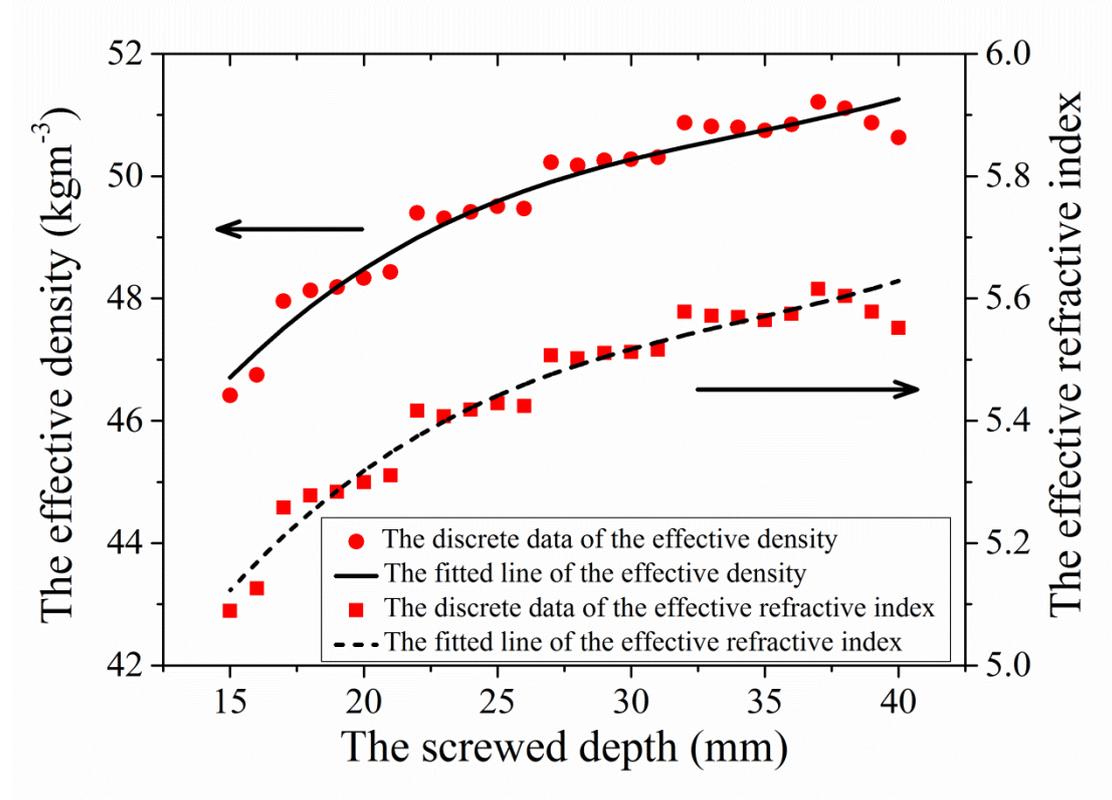

**Supplementary Figure 2** | The discrete and fitted results of the effective material parameters $n_{\text{eff}}$ and $\rho_{\text{eff}}$ associated with the screwed depth $L'$. The fitting curves are given by
$n_{\text{eff}}(L') = 3.37 \times 10^{-5}(L')^3 - 3.47 \times 10^{-3}(L')^2 + 0.129 L' + 3.85$ and
$\rho_{\text{eff}}(L') = 3.07 \times 10^{-4}(L')^3 - 3.17 \times 10^{-2}(L')^2 + 1.18 L' + 35.11$ (in kgm$^{-3}$).

# Supplementary Note 2. Additional results

The following two figures show the numerical simulation and experimental results for some other selected values of the frequency and focusing length.

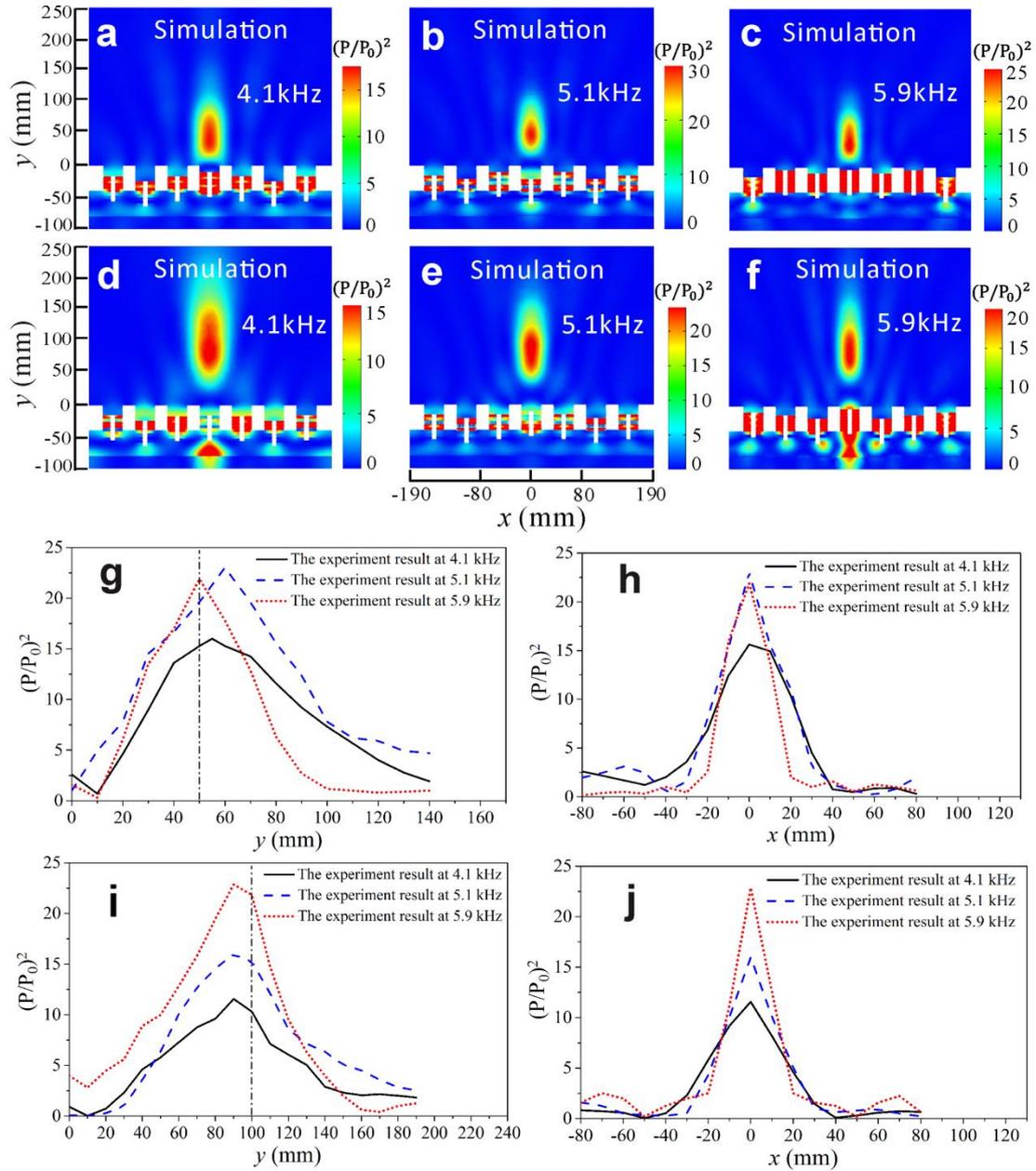

**Supplementary Figure 3│ The numerical and experimental focusing results for several selected frequencies when using the plan wave source.** (a)-(f) Numerical simulation results for the incident plane wave source at the frequencies of 4.1 kHz, 5.1 kHz and 5.9 kHz with the focusing lengths of 50mm (a-c) and 100mm (d-f). (g) and (h) The experimentally measured intensity fields at the three frequencies along the $y$- and $x$-directions at the focal spots for the focusing lengths of 50mm (g, h) and 100mm (i, j).

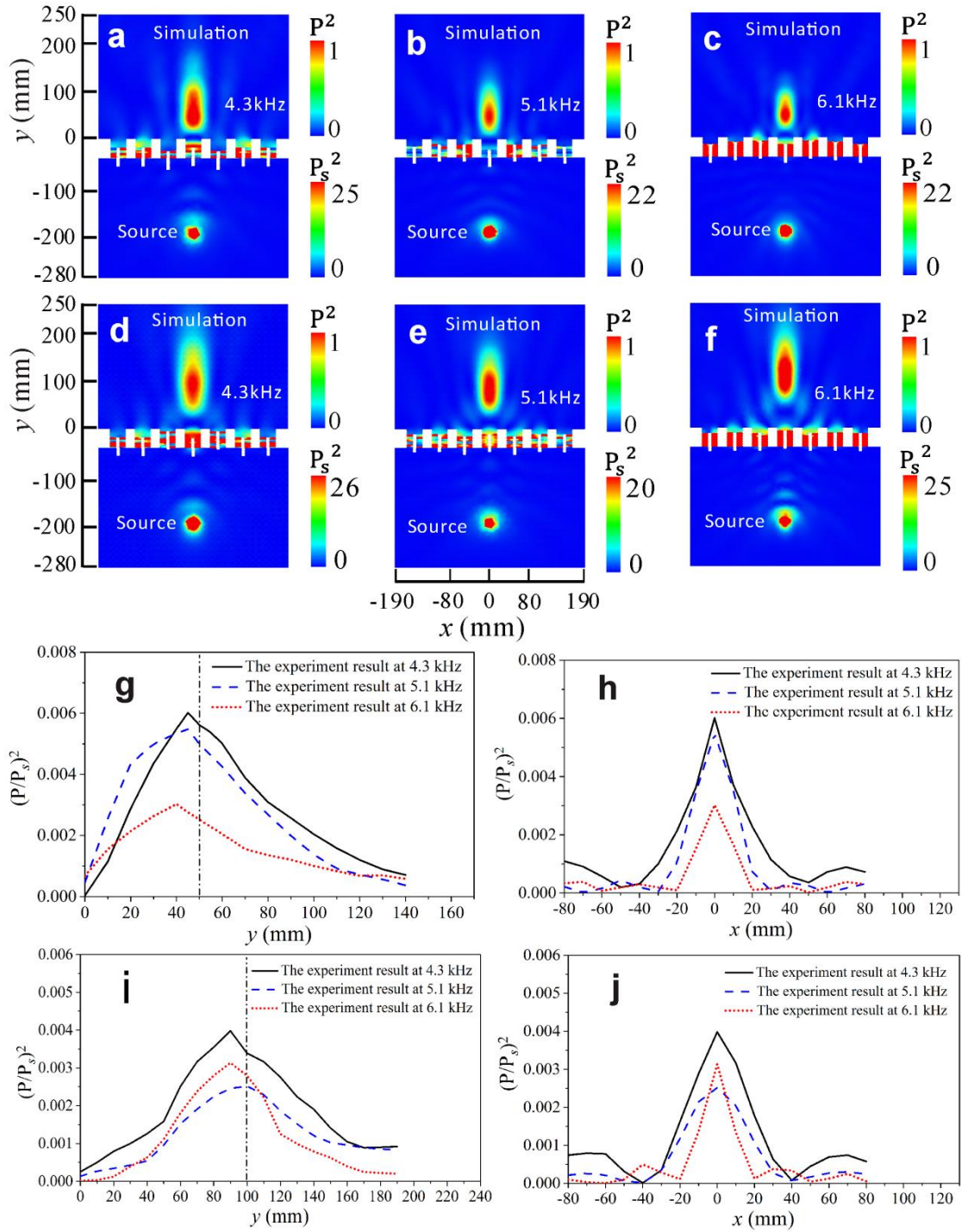

**Supplementary Figure 4 | The numerical and experimental focusing results for several selected frequencies when using the point source.** (a)-(f) Numerical simulation results for the point source ($y_s = 150$mm) at the frequencies of 4.3 kHz, 5.1 kHz and 6.1 kHz with the focusing lengths of 50mm (a-c) and 100mm (d-f). (g) and (h) The experimentally measured intensity fields at the three frequencies along the *y*- and *x*-directions at the focal spots for the focusing lengths of 50mm (g, h) and 100mm (i, j).

## Supplementary Note 3. Definition and calculation of $P_s$

In the numerical computation for the case of a point source, we suppose a time-harmonic acoustic pressure field at the source point and then calculate the surrounding sound field. However, in the experiment a semi-spherical surface sound source with the radius of 12.7mm is used. In order to compare the numerical results with the experimental results, we consider the semi-spherical surface sound source as the half of a radiated field from a point source. The radiated sound pressure from a point source is given by[2]

$$P(r) = j\frac{k_0 \rho_0 c_0}{4\pi r} Q_0 e^{-jk_0 r}, \qquad (S7)$$

where the common time-harmonic factor $e^{j\omega t}$ is suppressed; $j = \sqrt{-1}$; and $Q_0$ specifies the intensity of the source by stating the volume flow rate from the source. In the experiment, the pressure amplitude $P_s$ on the semi-spherical surface is measured and used in the normalization of the acoustic intensity field. From Eq. (S7), we have

$$P_s = \frac{1}{2}|P(r)| = \frac{k_0 \rho_0 c_0}{8\pi r} Q_0, \quad r = 12.7\text{mm}. \qquad (S8)$$

For comparison, the numerical results are also be normalized by $P_s$.

# Supplementary Note 4. Samples and flow chart of the experiment

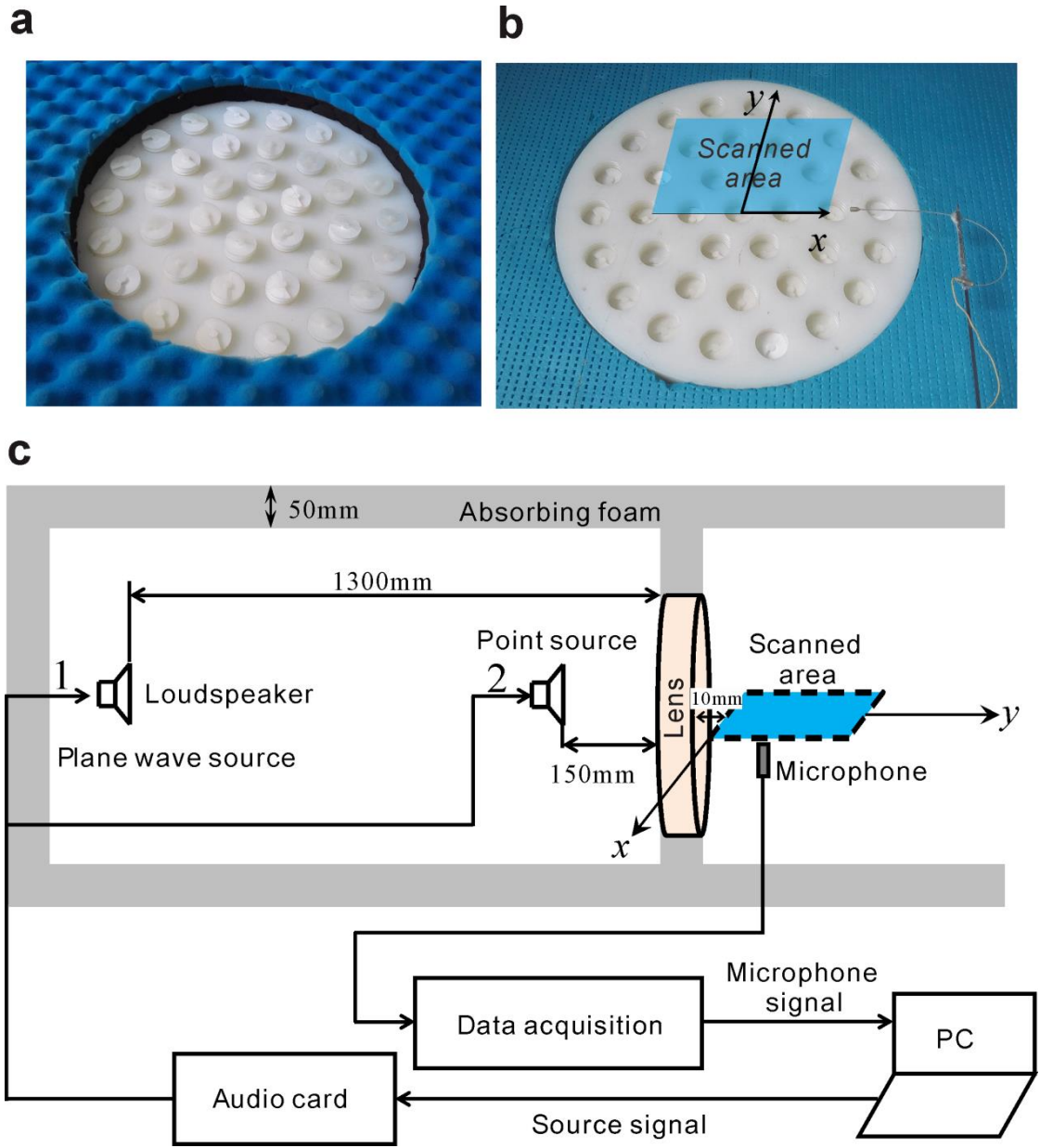

**Supplementary Figure 5 | The samples and flow chart of the experiment.** (a) The incident wave (front sample) side with the outer pillows of the HSM sample. (b) The back sample side where the focused wave field is measured; and (c) the flow chart of the experiment with the scanned area in the *x-y* plane and the absorbing foam surrounding the source and sample.

**Supplementary Table 1**│ The optimized screwed depth corresponding to variable frequencies (3.9kHz~6.3kHz), different wave sources (plane wave source and point source) and different focusing lengths (50mm and 100mm).

|  | f (kHz) | 3.9 | 4.1 | 4.3 | 4.5 | 4.7 | 4.9 | 5.1 | 5.3 | 5.5 | 5.7 | 5.9 | 6.1 | 6.3 |
|---|---|---|---|---|---|---|---|---|---|---|---|---|---|---|
| Plane wave source ($y_f$ = 50mm) | Layer 1 (mm) | 24.1 | 29.7 | 28.6 | 20.1 | 26.3 | 19.4 | 18.8 | 34.3 | 28 | 32.4 | 30.8 | 34 | 38 |
|  | Layer 2 (mm) | 17 | 23.5 | 22.5 | 33.1 | 20.6 | 31 | 30 | 29 | 22.8 | 27 | 26 | 16.2 | 33 |
|  | Layer 3 (mm) | 30.9 | 15.5 | 34.5 | 21.4 | 27 | 19.8 | 19.2 | 34.5 | 28 | 32 | 30.3 | 30.8 | 23.8 |
|  | Layer 4 (mm) | 17 | 23.2 | 22 | 28 | 19.4 | 25.6 | 24.6 | 23.6 | 32.8 | 21.6 | 31.9 | 34.4 | 38 |
| Plane wave source ($y_f$ = 100mm) | Layer 1 (mm) | 30.3 | 23.9 | 22.8 | 27.9 | 32.1 | 31.1 | 29.8 | 28.6 | 33 | 31.2 | 36 | 26 | 23.5 |
|  | Layer 2 (mm) | 24.8 | 23 | 22 | 26.5 | 27 | 26 | 25 | 24 | 17.2 | 27 | 21 | 24.9 | 33 |
|  | Layer 3 (mm) | 17 | 15.9 | 35 | 20.2 | 20.7 | 19.9 | 19 | 34.2 | 22.7 | 21.4 | 26.2 | 30 | 24.8 |
|  | Layer 4 (mm) | 31.1 | 24.4 | 22.7 | 27.5 | 27.8 | 26.3 | 24.9 | 23.8 | 28 | 26.2 | 30.9 | 34.5 | 28.8 |
| Point source ($y_f$ = 50mm) | Layer 1 (mm) | 23.4 | 29.8 | 22.7 | 22.2 | 25.4 | 19.9 | 19.1 | 34.9 | 22.2 | 21.6 | 25.1 | 25.8 | 37.4 |
|  | Layer 2 (mm) | 16.4 | 23 | 15.9 | 34.6 | 19.4 | 31.2 | 29.5 | 28.9 | 28.2 | 27.6 | 31 | 34 | 29 |
|  | Layer 3 (mm) | 22.5 | 28.6 | 21.9 | 21.1 | 20.3 | 31.8 | 30 | 29 | 28 | 27 | 30.3 | 30.6 | 27.5 |
|  | Layer 4 (mm) | 24 | 29.4 | 22 | 21 | 20 | 31 | 25.5 | 24.5 | 38.9 | 22.7 | 25.6 | 25.6 | 33.8 |
| Point source ($y_f$ = 100mm) | Layer 1 (mm) | 23.5 | 28.2 | 21.1 | 20.2 | 20.2 | 31.2 | 30.5 | 29.9 | 32 | 36.1 | 35.3 | 34.3 | 33.5 |
|  | Layer 2 (mm) | 16.8 | 22.9 | 35 | 33.4 | 32.7 | 25.6 | 24.9 | 28 | 27.6 | 31 | 30.3 | 29.4 | 28.7 |
|  | Layer 3 (mm) | 24.9 | 30 | 22 | 21 | 38.8 | 31 | 30 | 29.7 | 29 | 32 | 31 | 30 | 29 |
|  | Layer 4 (mm) | 31 | 15.1 | 22.5 | 21.4 | 39 | 30.8 | 29.5 | 29 | 28 | 30.5 | 30.1 | 29.2 | 24.9 |

## Supplementary references

1. J. F, Allard, N. Atalla, (2009) *Propagation of Sound in Porous Media: Modelling Sound Absorbing Materials* (John Wiley & Sons Ltd).
2. E. Skudrzyk, (1971) *The Foundations of Acoustics: Basic Mathematics and Basic Acoustics.* (Springer-Verlag New York Wien)